\documentclass[floatfix,showkeys,preprint]{revtex4-1}

\usepackage{amsmath}
\usepackage{amssymb}
\usepackage{graphicx}
\usepackage{pstricks}
\usepackage{pst-all}
\newcommand{\bvec}[1]{\ensuremath{\boldsymbol{#1}}}
\newcommand{\bmat}[1]{\ensuremath{{\bf #1}}}
\DeclareMathOperator{\tr}{Tr}

\begin{document}
\title{Efficient conversion from rotating matrix to rotation axis and angle by extending Rodrigues' formula}
\author{Kuo Kan Liang}
\email{kuokan.liang@gmail.com}
\affiliation{Research Center for Applied Sciences, Academia Sinica, Taipei 115, Taiwan}
\affiliation{Department of Biochemical Science and Technology, National Taiwan University, Taipei 106, Taiwan}
\affiliation{Genomics and Systems Biology Graduate Program, National Taiwan University and Academia Sinica, Taipei 106, Taiwan}

\begin{abstract}
In computational 3D geometric problems involving rotations, 
it is often that people have to convert back and forth between
a rotational matrix and a rotation described by an axis and a
corresponding angle. For this purpose,
Rodrigues' rotation formula is a very popular expression to use
because of its simplicity and efficiency.
Nevertheless, while converting a rotation matrix to an axis of rotation
and the rotation angle, there exists ambiguity.
Further judgement or even manual interference may be necessary 
in some situations.
An extension of the Rodrigues' formula helps to find the sine and
cosine values of the rotation angle with respect to a given rotation axis
is found and this simple extension may help to accelerate many 
applications.
\end{abstract}

\maketitle

\section{Introduction}
Description of rotation is a very important job in many applications.
For examples, in robotics, it is used in describing the state of
joints of robotic arms.
Also, for robots with vision, it is used to describe the relative
orientation of the robot with respect to a specific coordinate system.

For different computations in these applications, different forms
to describe rotations have their respective advantages.
Therefore, we may have to be able to convert a specific rotation
among different representations.
These representations may include, not exhaustively,
rotation matrix, Eular angles, rotation axis and angle, and quaternions.

Rodrigues' formula\cite{Rod1} is very popular for conversion between
rotation-matrix and rotation-axis/angle representations,
because the 3-dimensional expression of Rodrigues' theory is simple
and efficient for computation.
In this article, however, we point out that a very simple extension
of this well known formula can help to make its application even
more computationally favorable.
Interesting enough, this extremely simple extension cannot be found
in the literature, as far as the author could possibly find out.

\section{Review of the Rodrigues' rotation formula}
If a rotation axis $\boldsymbol n$ and a rotation angle $\theta$
with respect to $\boldsymbol n$ (in the counter-clockwise sense)
are given, one can construct the rotation matrix $\bmat{R}$
corresponding to this rotation.%
\footnote{In this article, a vector is typeset in italic bold symbol,
while a matrix is typeset in upright bold symbol,
if not explicitly indicated otherwise.}

Suppose that a vector $\bvec{v}$ is rotated by this rotation $\bmat{R}$
into a new vector $\bvec{u}$.
The part of $\bvec{v}$ parrallel to $\bvec{n}$ is never rotated,
and this part is expressed as
\begin{equation}
\bvec{v}_{\parallel}=\bvec{n}\left(\bvec{n}\cdot\bvec{v}\right)
\end{equation}
The part of $\bvec{v}$ perpendicular to $\bvec{n}$,
expressed as $\bvec{v}_{\perp}=\bvec{v}-\bvec{v}_{\parallel}$,
is the part actually rotated.
In other words, while indicating that
$\bvec{u}=\bmat{R}\bvec{v}$, what really happens during a rotation
is that $\bvec{u}_{\parallel}=\bvec{v}_{\parallel}$ and
$\bvec{u}_{\perp}=\bmat{R}\bvec{v}_{\perp}$.
In summary,
\begin{equation}
\bmat{R}\bvec{v}=\bvec{u}=\bvec{u}_{\parallel}+\bvec{u}_{\perp}
=\bmat{R}\bvec{v}_{\parallel}+\bmat{R}\bvec{v}_{\perp}
=\bvec{v}+\bmat{R}\bvec{v}_{\perp}
\end{equation}

To further explore the usage of this formula,
on may first want to derive the following vector identity:
\begin{equation}
\begin{aligned}
\bvec{n}\times\left(\bvec{n}\times\bvec{v}\right)&=
\varepsilon_{ijk}\hat{\bvec{e}}_in_j\varepsilon_{k\ell m}n_{\ell}v_m
=\varepsilon_{kij}\varepsilon_{k\ell m}\hat{\bvec{e}}_in_jn_{\ell}v_m
\\&=
\left(\delta_{i\ell}\delta_{jm}-\delta_{im}\delta_{j\ell}\right)
\hat{\bvec{e}}_in_jn_{\ell}v_m
=\hat{\bvec{e}}_in_in_jv_j-\hat{\bvec{e}}_iv_in_jn_j
\\&=
\bvec{n}\left(\bvec{n}\cdot\bvec{v}\right)-
\bvec{v}\left(\bvec{n}\cdot\bvec{n}\right)
=\bvec{n}\left(\bvec{n}\cdot\bvec{v}\right)-\bvec{v}
\end{aligned}
\end{equation}
where we have employed, at the same time, the Levi-Civita symbol
for representing cross product, the Kronecker delta function,
and the Einstein convention of summation over repeated indices.
It follows that
\begin{gather}
\bvec{v}_{\parallel}=\bvec{n}\left(\bvec{n}\cdot\bvec{v}\right)
=\bvec{v}+\bvec{n}\times\left(\bvec{n}\times\bvec{v}\right)\\
\bvec{v}_{\perp}=\bvec{v}-\bvec{v}_{\parallel}=
-\bvec{n}\times\left(\bvec{n}\times\bvec{v}\right)
\end{gather}

\begin{figure}
\begin{center}\begin{pspicture}(6,6)
\rput[lb](0,0){\includegraphics[clip,height=6cm]{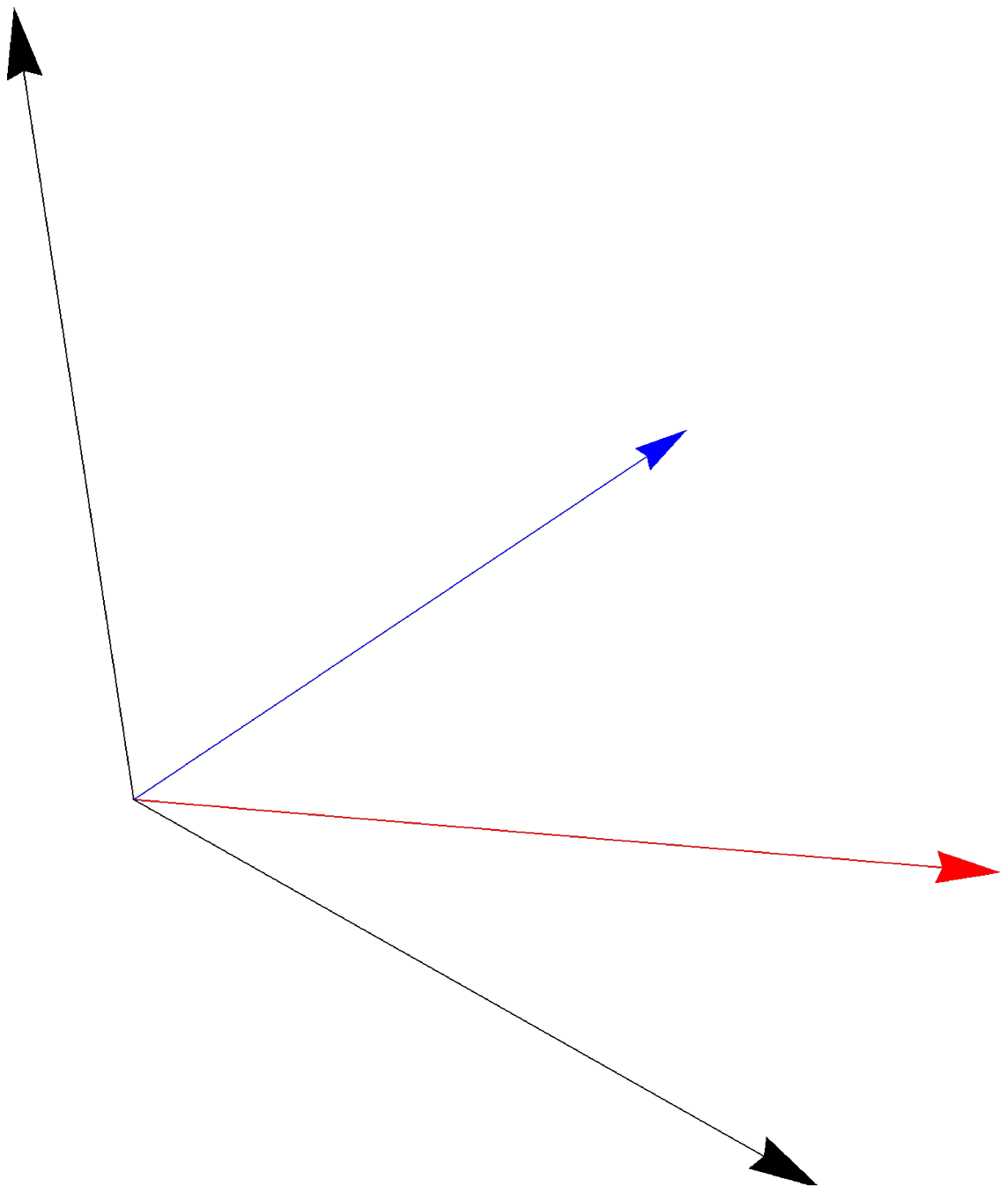}}
\rput(2.2,1.6){$\theta$}\rput(0.9,5.8){$\bvec{n}$}
\rput(4.7,0.2){$\bvec{v}_{\perp}$}
\rput(5.6,1.8){\textcolor{red}{$\bvec{u}_{\perp}$}}
\rput(4,4){\textcolor{blue}{$\bvec{n}\times\bvec{v}_{\perp}$}}
\end{pspicture}\end{center}
\caption{The spatial relation between $\bvec{n}$, $\bvec{v}$, 
and $\bvec{u}$.\label{f01}}
\end{figure}

According to Figure~\ref{f01}, 
if we are to express $\bvec{u}_{\perp}$ in terms of
$\bvec{n}$ and $\bvec{v}_{\perp}$, the part of $\bvec{u}_{\perp}$
parallel to $\bvec{v}_{\perp}$ is $\cos\theta\,\bvec{v}_{\perp}$,
and the part of $\bvec{u}_{\perp}$ perpendicular to $\bvec{v}_{\perp}$
should point in the direction of $\bvec{n}\times\bvec{v}_{\perp}$
with length $\left|\bvec{v}_{\perp}\right|\sin\theta$.
In other words, it is 
$\sin\theta\left(\bvec{n}\times\bvec{v}_{\perp}\right)$.
To sum up,
\begin{equation}
\begin{aligned}
\bvec{u}&=\bvec{u}_{\parallel}+\bvec{u}_{\perp}=
\bvec{v}_{\parallel}+\bvec{u}_{\perp}\\
&=\bvec{n}\left(\bvec{n}\cdot\bvec{v}\right)+
\cos\theta\,\bvec{v}_{\perp}
+\sin\theta\left(\bvec{n}\times\bvec{v}_{\perp}\right)\\
&=\bvec{n}\left(\bvec{n}\cdot\bvec{v}\right)+
\cos\theta\left(\bvec{v}
-\bvec{n}\left(\bvec{n}\cdot\bvec{v}\right)\right)
+\sin\theta\left(\bvec{n}\times
\left(\bvec{v}-\bvec{n}\left(\bvec{n}\cdot\bvec{v}\right)\right)\right)
\\
&=\left(1-\cos\theta\right)\bvec{n}\left(\bvec{n}\cdot\bvec{v}\right)+
\cos\theta\,\bvec{v}+\sin\theta\,\bvec{n}\times\bvec{v}
\end{aligned}
\label{erf0}
\end{equation}

The final expression of $\bvec{u}$ in Eq.~\eqref{erf0} is one form
of the Rodrigues' formula.
But since it involves specific $\bvec{v}$, 
it is formally not general enough.
One convention is to make use the antisymmetric tensor form of
the axial axis $\bvec{n}$ to express the formula.
Since we have already employed the Levi-Civita tensor in
previous equations, one simplest way to derive the form of 
the antisymmetric tensor is to make use again the Levi-Civita symbol.
The task is to express the cross product $\bvec{n}\times\bvec{v}$ as,
instead, the product of a matrix $\bmat{N}$ on $\bvec{v}$.

If we did find $\bmat{N}$ such that 
$\bmat{N}\bvec{v}=\bvec{n}\times\bvec{v}$, we have
\begin{equation}\label{edefN}
\left(\bmat{N}\bvec{v}\right)_i=\left(\bmat{N}\right)_{ik}v_k
=\left(\bvec{n}\times\bvec{v}\right)_i=
\varepsilon_{ijk}n_jv_k
\end{equation}
It becomes clear that
\begin{equation}
\left(\bmat{N}\right)_{ik}=\varepsilon_{ijk}n_j
\end{equation}
In explicit matrix form that means
\begin{equation}
\bmat{N}=\begin{pmatrix}
0 & -n_3 & n_2 \\ n_3 & 0 & -n_1 \\ -n_2 & n_1 & 0
\end{pmatrix}
\end{equation}
The remaining expression, $\bvec{n}\left(\bvec{n}\cdot\bvec{v}\right)$,
can also be easily represented by the product of a matrix with 
$\bvec{v}$:
\begin{equation}
\bvec{n}\left(\bvec{n}\cdot\bvec{v}\right)=
\begin{pmatrix}\uparrow\\ \bvec{n}\\ \downarrow\end{pmatrix}
\left(
\begin{pmatrix}\leftarrow & \bvec{n} & \rightarrow\end{pmatrix}
\begin{pmatrix}\uparrow\\ \bvec{v}\\ \downarrow\end{pmatrix}
\right)=
\left(
\begin{pmatrix}\uparrow\\ \bvec{n}\\ \downarrow\end{pmatrix}
\begin{pmatrix}\leftarrow & \bvec{n} & \rightarrow\end{pmatrix}
\right)
\begin{pmatrix}\uparrow\\ \bvec{v}\\ \downarrow\end{pmatrix}
=\left(\bvec{n}\bvec{n}^{\rm T}\right)\bvec{v}
\end{equation}
In summary,
\begin{equation}
\bmat{R}\bvec{v}=\bvec{u}=
\cos\theta\,\bvec{v}+\left(1-\cos\theta\right)
\left(\bvec{n}\bvec{n}^{\rm T}\right)\bvec{v}+
\sin\theta\,\bmat{N}\bvec{v}
\end{equation}
and the general form of the Rodrigues' formula becomes
\begin{equation}
\bmat{R}=\cos\theta\,\bmat{1}+
\left(1-\cos\theta\right)\bvec{n}\bvec{n}^{\rm T}
+\sin\theta\,\bmat{N}
\end{equation}
where $\bmat{1}$ is the unity matrix.
This formula, however, can be further simplified to only use
matrices $\bmat{1}$ and $\bmat{N}$ throughout.
Notice that 
$\bvec{n}\left(\bvec{n}\cdot\bvec{v}\right)=
\bvec{v}+\bvec{n}\times\left(\bvec{n}\times\bvec{v}\right)$,
we have
$\bvec{n}\bvec{n}^{\rm T}=\bmat{1}+\bmat{N}\bmat{N}$.
Therefore
\begin{equation}\label{erf1}
\bmat{R}=\cos\theta\,\bmat{1}+
\left(1-\cos\theta\right)\left(\bmat{1}+\bmat{N}\bmat{N}\right)+
\sin\theta\,\bmat{N}=\bmat{1}+\sin\theta\,\bmat{N}+
\left(1-\cos\theta\right)\bmat{N}\bmat{N}
\end{equation}
This final expression in Eq.~\eqref{erf1} is not only neat 
but is more convenient for deriving important equations
we shall present later.

If one is given the rotation matrix $\bmat{R}$ and has to find out
the rotation matrix $\bvec{n}$ and rotation angle $\theta$,
it can be done in two steps.
First, since $\bvec{n}$ is invariant under rotation by $\bmat{R}$,
it can be thought that $\bvec{n}$ satisfies the eigenvalue equation
\begin{equation}\label{ees0}
\bmat{R}\bvec{n}=1\,\bvec{n}
\end{equation}
In this article, we shall not discuss how to solve this equation.

In the second step, one tries to find the value of $\theta$.
Conventionally, one takes the trace of $\bmat{R}$.
Since $\tr\left(\bmat{1}\right)=3$, 
$\tr\left(\bvec{n}\bvec{n}^{\rm T}\right)=1$,
$\tr\left(\bmat{N}\right)=0$, and therefore
$\tr\left(\bmat{N}\bmat{N}\right)=1-3=-2$, we can find that
\begin{equation}\label{ecos0}
\tr\left(\bmat{R}\right)=3-2\left(1-\cos\theta\right)=2\cos\theta+1
\Rightarrow\cos\theta=\frac{\tr\left(\bmat{R}\right)-1}{2}
\end{equation}
Therefore,
\begin{equation}\label{eth0}
\theta=\cos^{-1}\left(\frac{\tr\left(\bmat{R}\right)-1}{2}\right)
\end{equation}

Conventionally, $\bvec{n}$ and $\theta$ are found out in the
above two steps. But there are severe problems.

It is clear that there are typically two nontrivial eigenvectors found 
by solving Eq.~\eqref{ees0}, and they are anti-parallel to each other.
Meanwhile, the $\theta$ value obtain from Eq.~\eqref{eth0} is also
not unique. There are typically two $\theta$ values satisfying the
same equation and they differ from each other by a sign. 
This is a well known property of the trigonometric functions.
In other words, one will get four different possible solutions
of the following forms:
$\left(\bvec{n},\theta\right)$, $\left(-\bvec{n},\theta\right)$,
$\left(\bvec{n},-\theta\right)$, and
$\left(-\bvec{n},-\theta\right)$.
Only two of them, namely, $\left(\bvec{n},\theta\right)$ and
$\left(-\bvec{n},-\theta\right)$, represent the correct rotation.

In the following Section, we shall derive another equation for
accessing the value of $\sin\theta$, 
given a preferred value of $\bvec{n}$.
This new addition to the conventional Rodrigues' formula
helps to pick the better $\bvec{n}$ and $\theta$ values.

\section{Extension}
The eigenvalue system in Eq.~\eqref{ees0} must be solved anyway.
Suppose that we have already pick one of the solution $\bvec{n}$,
the antisymmetric tensor $\bmat{N}$ can immediately be constructed.
Then, we multiply Eq.~\eqref{erf1} throughout by $\bmat{N}$
and find that
\begin{equation}
\bmat{N}\bmat{R}=\bmat{N}+\sin\theta\,\bmat{N}\bmat{N}+
\left(1-\cos\theta\right)\bmat{N}\bmat{N}\bmat{N}
\end{equation}
Now it can be reasoned that since 
$\bmat{N}\bmat{N}=\bvec{n}\bvec{n}^{\rm T}-\bmat{1}$
and $\bmat{N}\bvec{n}=0$, 
we have $\bmat{N}\bmat{N}\bmat{N}=-\bmat{N}$.
Therefore
\begin{equation}
\bmat{N}\bmat{R}=\bmat{N}+\sin\theta\,\bmat{N}\bmat{N}-
\left(1-\cos\theta\right)\bmat{N}=
\cos\theta\,\bmat{N}+\sin\theta\,\bmat{N}\bmat{N}
\end{equation}
Thus,
\begin{equation}\label{esin0}
\tr\left(\bmat{N}\bmat{R}\right)=-2\sin\theta\Rightarrow
\sin\theta=-\frac{\tr\left({\bmat{N}\bmat{R}}\right)}{2}
\end{equation}

As we mentioned in the previous Section, the value of $\theta$
cannot be uniquely determined with only the value of 
$\cos\theta$ from Eq.~\eqref{ecos0}, even when $\bvec{n}$
is already given.
Notice that equations \eqref{ecos0} and \eqref{eth0}
do not depend on $\bvec{n}$.
But together with the given axis $\bvec{n}$,
plus the value of $\sin\theta$ from Eq.~\eqref{esin0},
the value of $\theta$ can be computed by the two-argument
inverse-tangent function available in any computational system.

Surprisingly, this simple relation could not be found in the
literature, to the knowledge of the author.

\section{Numerical example}
In this Section we use a numerical example to demonstrate the
problems we mentioned in earlier Sections and show that
the extension proposed in the previous Section does help to 
simplify the jobs.

Mathematica codes were used because analytic expressions and
numerical results can conveniently be shown in one place.

First, suppose that a rotation axis $\bvec{n}$ and rotation
angle $\theta$ is given.
According to the discussion on Rodrigues' formula,
one can first construct the antisymmetric tensor $\bmat{N}$,
and then the rotation matrix $\bmat{R}$.

In Mathematica, there are more systematic ways of
constructing the antisymmetric tensor with the help of
Levi-Civita tensor.
However, this approach is beyond the scope of this article.
Therefore, instead, a more cumbersome statement reflecting
Eq.~\eqref{edefN} is used to construct $\bmat{N}$ from $\bvec{n}$:

\vspace{1em}\includegraphics[width=0.96\textwidth]{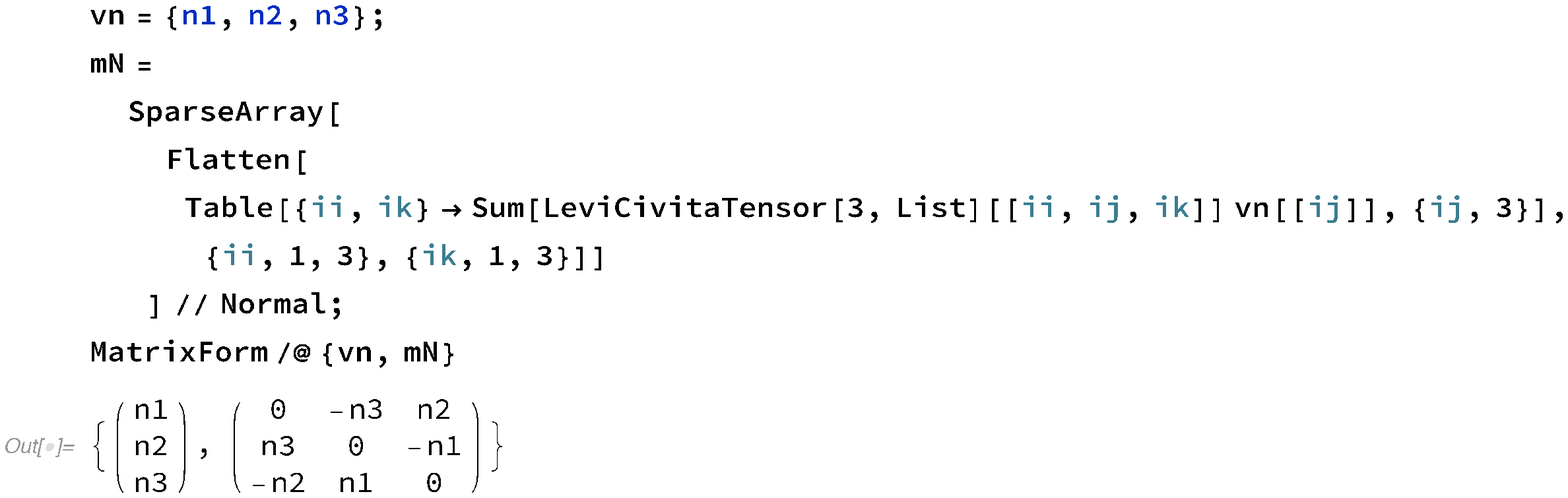}

Then we can use the final result of Eq.~\eqref{erf1} to construct
$\bmat{R}$:

\vspace{1em}\includegraphics[width=0.96\textwidth]{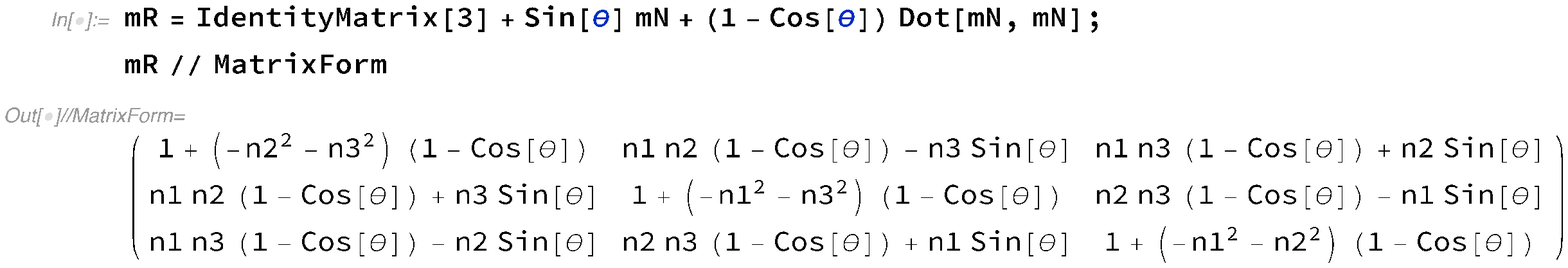}

To show the usage of the Rodrigues' formula and its extension,
now we try to recover $\bvec{n}$ and $\theta$ from the 
rotation matrix $\bmat{R}$ just constructed.
As a numerical example, let $\bvec{n}$ be 
$\left(1/\sqrt2,1/\sqrt2,0\right)$ and $\theta=\pi/6$.
Then we solve the eigensystem problem in Eq.~\eqref{ees0}:

\vspace{1em}\includegraphics[width=0.96\textwidth]{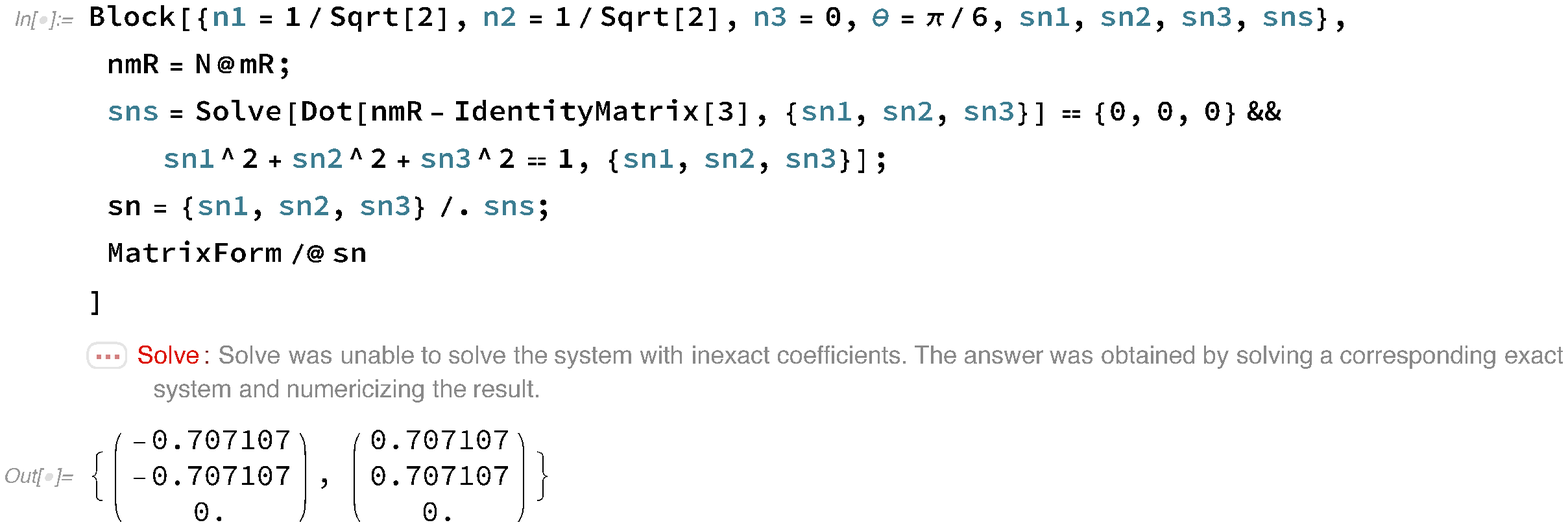}

Beside the orginal $\bvec{n}$ we used, $-\bvec{n}$ is also listed as
a solution, as expected. Now, let us try to find $\theta$ for
both solutions.

To demostrate the problem of the conventional Rodrigues' formula,
we use Eq.~\eqref{eth0} to find $\theta$:

\vspace{1em}\includegraphics[width=0.3\textwidth]{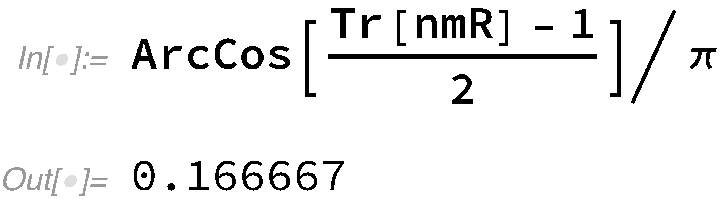}

This is indeed $\pi/6$. Nevertheless, if we have no prior idea whether
this angle is related to 
$\bvec{n}=\left(1/\sqrt{2},1/\sqrt{2},0\right)$ or $-\bvec{n}$,
we may get wrong answer.
Even if we know this problem, however, there is nothing that we can do
because we could not use the information about $\bvec{n}$ when
calculating $\theta$.

Using the extended Rodrigues' formula, however, is a new story.
In the following code, we demonstrate the result of finding all
correct answers with the information we obtained from the previous
piece of codes:

\vspace{1em}\includegraphics[width=0.7\textwidth]{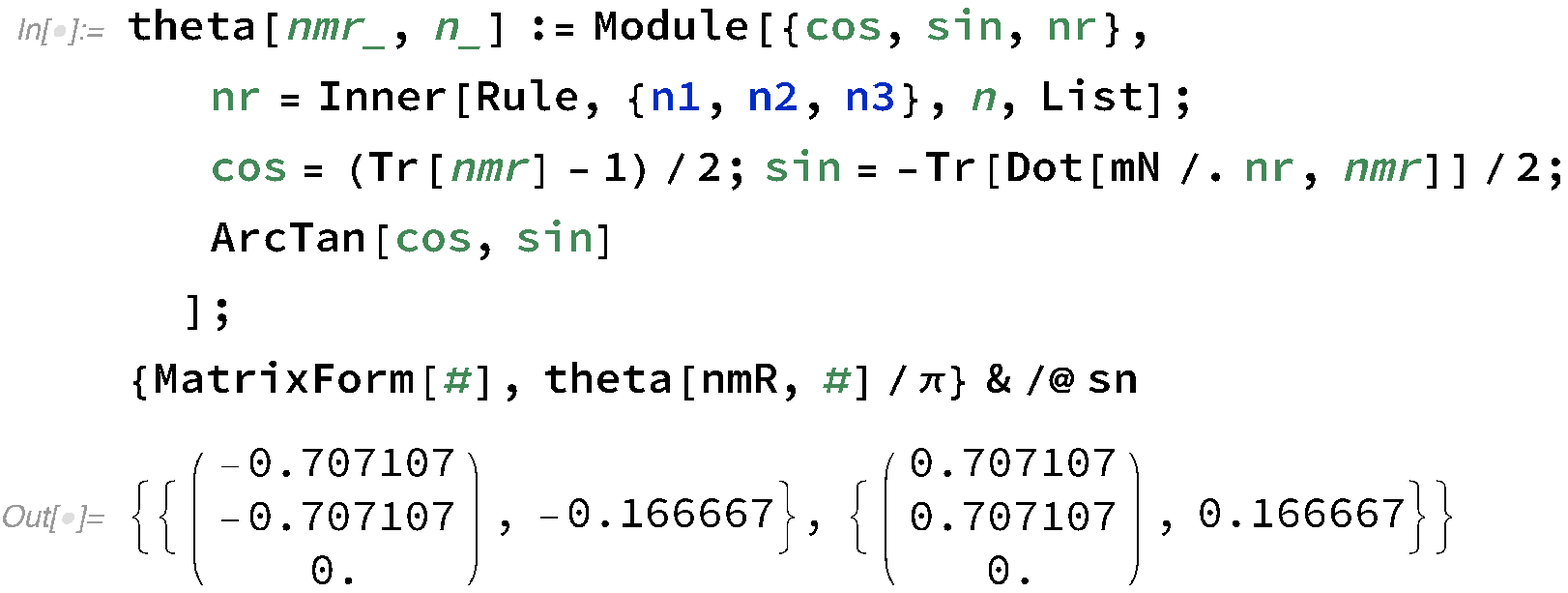}

Depending on the application, there will be different strategies of
choosing the more appropriate $\bvec{n}$.
For example, the rotation axis may be changing gradually and
we hope that the new axis we computed to be the one closer to
the previous one used. In that case we would choose the one with
the largest inner product with the previous axis.
In any case, as long as the $\bvec{n}$ is uniquely determined,
the correct rotation angle $\theta$ can be correctly found,
without requiring human intervention or complex codes for judgment.

\section{Summary}

For people using Rodrigues' formula extensively in their codes,
in any case the matrices $\bmat{N}$ and $\bmat{R}$ will be
computed. Therefore, using the equations \eqref{ecos0} and \eqref{esin0}
to find $\cos\theta$ and $\sin\theta$ and then use inverse tangent
function to find $\theta$ is of very little burden to the program.
However, it helps a lot to reduce the need of judging whether the
angle found indeed correspond to the axis one desires to use.
It is hoped that this extension, contributed by the author,
can help the community to simplify their codes and make the codes
clearer, more correct, and faster.

\begin{acknowledgments}
This work is supported by the internal grand of the Research Center
for Applied Sciences (RCAS) of Academia Sinica (AS), Taiwan.
\end{acknowledgments}

\bibliography{rr}

\end{document}